\documentclass[10pt,twocolumn]{article}

\usepackage[utf8]{inputenc}
\usepackage{amsmath, amssymb}
\usepackage{graphicx}
\usepackage[colorlinks]{hyperref}
\usepackage{geometry}
\usepackage{xcolor}
\usepackage{microtype}

\numberwithin{equation}{section}
\newcommand{\argmin}{\operatorname*{argmin}}

\usepackage{booktabs}
\usepackage{caption}
\begin{document}

\title{Model Predictive Control For Trade Execution}
\author{
    Thomas P. McAuliffe\textsuperscript{1}\footnotemark[1] ,
    Samuel Liew\textsuperscript{1},
    Yuchao Li\textsuperscript{2},
    Andrey Ushenin\textsuperscript{1}, \\
    Chihang Wang\textsuperscript{1},
    Alexandros Tasos\textsuperscript{1},
    Jack Pearce\textsuperscript{1},
    Dimitris Tasoulis\textsuperscript{1}, \\
    Dimitri P. Bertsekas\textsuperscript{1, 2, 3},
    Theodoros Tsagaris\textsuperscript{1} \\[0.3cm]
    \small \textsuperscript{1}Bayforest Technologies, London, UK \\
    \small \textsuperscript{2}Arizona State University, Tempe, AZ, USA \\
    \small \textsuperscript{3} Massachusetts Institute of Technology, Cambridge, MA, USA
}

\date{\today}

\twocolumn[
\begin{@twocolumnfalse}
\maketitle
\begin{abstract}
\noindent We address the problem of executing large client orders in continuous double-auction markets under time and liquidity constraints. We propose a model predictive control (MPC) framework that balances three competing objectives: order completion, market impact, and opportunity cost. Our algorithm is guided by a trading schedule (such as time-weighted average price or volume-weighted average price) but allows for deviations to reduce the expected execution cost, with due regard to risk.

Our MPC algorithm executes the order progressively, and at each decision step it solves a fast quadratic program that trades off expected transaction cost against schedule deviation, while incorporating a residual cost term derived from a simple base policy. Approximate schedule adherence is maintained through explicit bounds, while variance constraints on deviation provide direct risk control. The resulting system is modular, data-driven, and suitable for deployment in production trading infrastructure.

Using six months of NASDAQ `level 3' data and simulated orders, we show that our MPC approach reduces schedule shortfall by approximately 40-50\% relative to spread-crossing benchmarks and achieves significant reductions in slippage. Moreover, augmenting the base policy with predictive price information further enhances performance, highlighting the framework's flexibility for integration with forecasting components.
\end{abstract}

\vspace{0.5cm}
\end{@twocolumnfalse}
]

{
  \renewcommand{\thefootnote}{\fnsymbol{footnote}}
  \footnotetext[1]{Email: \texttt{tom.mcauliffe@bayforest.ai}}
}

\section{Introduction}
\label{sec:introduction}
We consider the design of an algorithm to execute a client's order in a continuous double-auction market. This is a trading mechanism in which both buyers and sellers submit bids and offers simultaneously, and transactions occur whenever the bids and offers are `marketable' (when a buyer's bid equals or exceeds a seller's ask). It is the dominant structure of modern exchanges and trading platforms.

The problem has received significant attention by both the professional and academic communities. The objective is to balance three competing priorities: completing the order within a given time period, minimizing purchase cost penalties due to market impact, and minimizing opportunity cost from subsequent price improvement. We illustrate these trade-offs with two contrasting policies, assuming that a fixed buy order is to be executed within a given time window:

(1) One possibility is to send a market order for the full quantity to the double-auction market. This means that we will immediately trade at the best price offered. In the absence of sufficient offered quantity at the best quote, the order `sweeps the book,' executing piecemeal across progressively worse price levels (for a buy: increasingly higher asks). The book subsequently refills as other participants and market makers re-quote, but the strong demand induces price impact. The order completes almost instantly (assuming there is sufficient volume currently available), but at the cost of maximal market impact and an average execution price materially worse than the arrival price due to the sweeping.

(2) The opposite extreme is to slice the client order into tiny clips and post highly passive limit orders (e.g. offer to buy at the current bid or slightly below) throughout the given time window. Then the market price impact is small, and the execution price tends to track the contemporaneous market average. However, the completion risk is high: orders may never trade, queue position can be lost due to cancels/requotes, and favorable price moves may not be properly exploited. 

In practice, good execution policies lie between these extremes, balancing completion, impact, and slippage.\footnote{Slippage generally refers to a measure of the difference between an order’s execution price and a specified benchmark price. Several types of slippage relevant to our trade execution setting will be discussed later; see Section \ref{subsec:metrics}.} There is a rich academic literature on the topic, but much of it includes assumptions that violate practical constraints. In this paper, we develop a realistic, production-grade algorithm that balances theoretical considerations with practical concerns and constraints. Because of its speed and flexibility, our algorithm can be deployed in a scalable, modular environment that coordinates orders across venues, order types, and brokers.

\subsection{Current Practices Review}
\label{subsec:current_practices}

Broadly speaking, the industry's most common approach to order execution aims to minimize deviation from a benchmark volume-weighted average price (VWAP), as introduced by Berkowitz \textit{et al.} \cite{berkowitz1988total}. In practice, such algorithms follow a trading schedule that tracks the market's aggregate executed volume over the trading day.

The intraday volume profile exhibits a characteristic U-shape: trading activity is highest at the market open and close \cite{bennett2012measuring}. This strategy aligns with market intuition: executing more when liquidity is abundant mitigates market impact (known to scale with participation rate) \cite{cont2014price}. Executing brokers and electronic trading platforms implement variations of this VWAP-based approach \cite{morganstanley2024-us-cash-equity-faq, jpmorgan2019-fx-ml, ubs2019-fx-algos}.

Bertsimas and Lo \cite{bertsimas1998optimal} formulated the optimal execution problem as a dynamic programming (DP) problem minimizing market impact costs. Under a linear impact model, they showed that the solution is a time-weighted average price (TWAP) strategy, now a standard baseline among practitioners. Almgren and Chriss \cite{almgren2000optimal} extended this framework by introducing a mean-variance formulation that penalizes cost uncertainty. In analogy with modern portfolio theory, they derived an efficient frontier of execution paths that minimize arrival slippage for a given level of risk. Both formulations determine a static schedule prior to trading, based on modeled price, impact dynamics, and a specified risk aversion parameter.

Cartea and Jaimungal \cite{cartea2015optimal} further enhanced these models by allowing a mixture of market and limit orders. Their algorithm executes passively when running ahead of schedule, thereby earning part of the spread, and resorts to market orders to catch up when behind schedule. This structure introduces an online, rule-based decision component, bridging theoretical models and practical execution logic.

Busetti and Boyd \cite{busseti2015volume} studied optimal execution under a VWAP benchmark for a risk-averse broker. They model slippage as a mean-variance objective incorporating quadratic transaction costs, and propose both a static execution strategy, computed before trading, and a dynamic strategy that adapts to information about volumes revealed during the day. The dynamic method embeds the problem within a linear-quadratic stochastic control framework and employs DP to address uncertainty in total daily market volume. Using real NYSE data and a log-normal model of intraday volumes, they show that the dynamic strategy reduces both VWAP deviation and execution costs compared with the standard static solution.

Reinforcement learning (RL) is a natural framework for modelling online sequential decision making under uncertainty. A substantial body of work has explored RL-based execution strategies \cite{nevmyvaka2006reinforcement, moallemi2022reinforcement, li2022hierarchical}. Nevmyvaka \textit{et al.} \cite{nevmyvaka2006reinforcement} developed one of the first empirical RL systems trained directly on NASDAQ limit order data. Their agent controls order aggressiveness $u_t$, posting passively for $u_t < 0$ and crossing the spread for $u_t > 0$. Hendricks and Wilcox \cite{hendricks2014reinforcement} proposed a formulation that maps their action space to the fraction of the Almgren-Chriss schedule executed in each interval. The intuition is that overshooting the schedule may be advantageous when volume is high and spreads are tight. Their approach uses Q-learning to effect this adjustment dynamically.

Moallemi and Wang \cite{moallemi2022reinforcement} focused on optimizing child orders by modeling the current stage cost around short duration price trajectory forecasts. They considered several approximations: (1) directly forecasting the sum of price returns, (2) a temporal-difference (TD) learning variant that exploits intermediate returns, and (3) a continuation-value approach estimating the benefit of deferring execution. For (2) and (3), they employed double deep Q-networks (DDQN).

Li \textit{et al.} \cite{li2022hierarchical} proposed the separation of tasks across three levels: (1) macro-level estimation of daily volume profiles, (2) meta-level selection of intermediate order quantities for a given tranche of the parent, and (3) microstructural-level submission of individual child orders. Such task separation, often termed hierarchical reinforcement learning \cite{dayan1993feudal, sutton1999between}, tends to improve scalability and interpretability.

There has also been interest in the use of model predictive control (MPC) methods for trade execution and portfolio optimization. For example, Clinet \textit{et al.} \cite{clinet2021optimal} proposed a MPC method by modeling a trading execution problem using a linear state equation, quadratic cost, and additional positivity constraints. Plessen and Bemporad \cite{plessen2017stock} studied the performance of multiple MPC methods designed for stock trading under the assumption of proportional transaction costs.  Other related MPC methods can be found in the references quoted in these papers. Note that MPC is closely related to RL. In fact, some of the most reliable RL methods can be viewed as a form of MPC; see \cite{bertsekas2024model}, \cite{bertsekas2025course}. 

The methodology of this paper bears a conceptual relation to the literature cited above. However, our framework allows for the flexible use of current and historical market data, includes multiple modular components that can be designed independently, and allows for fast execution of a variety of trading actions, as we will discuss shortly.

\subsection{A Summary of our Approach}
\label{subsec:our_approach}

Our work aims to develop a policy for placing orders at each of $T$ time periods. It balances a high rate of completion, a small expected mean and variance of trading cost, and relatively small deviation from a schedule such as VWAP. At each time period the policy submits multiple orders at varying prices based on an MPC optimization. It uses a cost approximation for future stages, and applies constraints on the deviation from the schedule. A detailed mathematical formulation will be given later. The optimization is very fast, and allows a large number of orders to be placed simultaneously at different price levels. Our approach takes into account stochastic uncertainties about the execution of the placed orders at the current time, and about the market price at future times.

Our algorithmic design is consistent with our view that a scalable, practical trading algorithm should satisfy the following constraints:

\begin{enumerate}
    \item \textbf{Data driven.} We do not wish to make many assumptions about market dynamics, and instead prefer to measure and respond online. Similar to Hendricks and Wilcox \cite{hendricks2014reinforcement} and others, we should incorporate as much recent market state information into the system as possible to improve online decision making. Various components of our framework naturally lend themselves to data-driven learning.

    \item \textbf{Fast decision times.} In a live trading environment we are managing hundreds to thousands of simultaneous orders. If each intraday decision requires many milliseconds, state information will be very stale by the time of actually taking an action.

	\item \textbf{Well separated concerns.} Similar to Li \textit{et al.} \cite{li2022hierarchical}, we argue that a robust system should be composed of single-responsibility services with well-defined key performance indicators. Such separation enables rigorous testing, isolated improvement and introspection of individual components, and better parallelization across researchers and developers.

	\item \textbf{Rich action space.} The literature exclusively focuses on action spaces defining simple limit and market orders (which interact with a simulated exchange), the most advanced of which use order `level 3' (L3) data to generate fills. In practice, there is a huge variety of order types and parameterizations, broker algorithms, exchanges, and alternative trading systems available as liquidity sources. We wish to fully account for these possibilities in our implemented algorithm.

\end{enumerate}

\section{Problem Formulation and the MPC Methodology}
\label{sec:problem_formulation}

In this section, we provide a high level summary of the MPC algorithmic framework as applied to our problem. It involves a stochastic discrete-time system and sequential decision making over $T$ time periods (see the textbook by Bertsekas \cite{bertsekas2025course} and references quoted there).

Denoting time by $t$, the system involves a state (denoted by $x_t$\footnote{This state contains both market (prices, volatility, etc) and order level (executed quantity, schedule, etc) data.}), a decision/control (denoted by $u_t$), a random quantity that models uncertainty (denoted by $w_t$), and a function $f_t$, which governs the evolution of the system's state:
$$x_{t+1} = f_t(x_t, u_t, w_t), \quad t = 0, 1, \dots, T - 1.$$
The control $u_t$ is to be selected from a given constraint set $U_t(x_t)$ that depends on the state $x_t$. The probability distribution of $w_t$ is given and may depend on $(x_t, u_t)$. The transition from $x_t$ to $x_{t+1}$ incurs a cost $g_t(x_t, u_t, w_t)$, and there is an additional cost $g_T(x_T)$ at the terminal time $T$ to account for the terminal state $x_T$. 

We aim to minimize the expected value of the total cost
$$g_T(x_T) + \sum_{t=0}^{T-1} g_t(x_t, u_t, w_t)$$
with an appropriate choice of each $u_t \in U_t(x_t)$ as a function of $x_t$.

\subsection{The Exact DP Algorithm}
\label{subsec:exact_dp}
The optimal solution can be found in principle by the DP algorithm. The exact version of DP computes for all $x_t$ and $t$, the scalar $J^*_t(x_t)$, which is the optimal cost starting at state $x_t$ and going to the end of the horizon $T$. Then, an optimal decision at time $t$ and state $x_t$ is obtained from the minimization
\begin{equation}
    u^*_t \in \argmin_{u_t \in U_t(x_t)} E \Big\{ g_t(x_t, u_t, w_t) + J^*_{t+1}(f_t(x_t, u_t, w_t)) \Big\}, \label{eq:exact_dp}
\end{equation}
where $E\{\cdot\}$ denotes expected value with respect to the probability distribution of $w_t$.

This encodes the classical DP principle: At each $t$ we should minimize the sum of the cost at the current time $t$ plus the future costs, assuming that we will make optimal choices at the future times $t + 1, \dots, T - 1$.

\subsection{The Approximate DP Algorithm}
\label{subsec:approx_dp}
Since computing the optimal cost functions $J^*_t$ is intractable for our problem, approximate DP/RL replaces $J^*_{t+1}$ with an approximation $\tilde{J}_{t+1}$ in Eq.~\eqref{eq:exact_dp}, and computes an approximately optimal decision $\tilde{u}_t$ according to
\begin{equation}
    \tilde{u}_t \in \argmin_{u_t \in U_t(x_t)} E\Big\{ g_t(x_t, u_t, w_t) + \tilde{J}_{t+1}(f_t(x_t, u_t, w_t)) \Big\}. \label{eq:approx_dp}
\end{equation}
This is the MPC method with one-step lookahead minimization. [A multistep version of MPC, involves minimization of the cost of multiple stages, say $k$, followed by $\tilde{J}_{t+k}(x_{t+k})$. We will not consider it here, although it is an interesting possibility for future work.]

The method is also referred to as {\it approximation in value space}, and is one of the most effective and reliable RL methods. Naturally, the computation of $\tilde{J}_{t+1}$ is an important issue. In our case it will be done with a form of the rollout algorithm, whereby $\tilde{J}_{t+1}$ approximates the cost function corresponding to some policy, starting at time $t + 1$.

\section{Trade Execution Model}
\label{sec:trade_execution_model}
We will now describe our MPC method \eqref{eq:approx_dp} as applied to trade execution. At each time $t$, it solves a quadratic programming problem of the form:
\begin{equation*}
    \min_{u_t \in U_t(x_t)} E \left[ 
    \begin{aligned}
        \text{Current Stage Trading Cost} \\
        + \text{Schedule Deviation Penalty} \\
        + \text{Future Cost Approximation}
    \end{aligned}
    \right]
\end{equation*}
The first and second terms above correspond to $g_t$ of Eq.~\eqref{eq:approx_dp}, while the third term corresponds to $\tilde{J}_{t+1}$.

For simplicity, we will assume in this section a uniform time discretization, i.e., that the time horizon is divided into $T$ equally spaced time steps, and that orders are submitted, filled and cancelled at the times $t = 0, 1, \dots, T - 1$. However, our MPC methodology applies to the more general case where the duration of an order may be longer or shorter than one unit. Theoretically, this involves no major difficulty, and can be done by using a more complicated definition of the state $x_t$, which additionally encodes the backlog of orders that have not been processed by the end of a time period; see, cf. \cite[Section 1.6]{bertsekas2025course}. Indeed, our implementation, described in Section~\ref{sec:model_implementation}, can be modified to account for orders of variable duration.

Using notation to be introduced shortly and the uniform time discretization assumption, the preceding minimization takes the form
\begin{equation}
\label{eq:trading_problem}
    \begin{aligned}
	\min_{u_t \in U_t(x_t)}  
	&\underbrace{(c_t \circ \pi_t)' u_t}_{\text{Trading Cost}} + \underbrace{\gamma(q_t + \pi_t' u_t - s_{t+1})^2}_{\text{Schedule Deviation}} \\
	&+ \underbrace{\xi_t (Q - q_t - \pi_t' u_t)}_{\text{Future Cost}}
\end{aligned}
\end{equation}
where $\circ$ denotes componentwise vector product, and a prime denotes vector transpose. Our notation is as follows:
\begin{itemize}
	\item $T$: duration of the parent order 
	\item $Q$: face quantity of the parent order
	\item $q_t$: the accumulated position at time $t$ (the total quantity of the orders that have been filled up to $t$)
	\item $u_t$: The vector of distinct order sizes that are placed at time $t$ (the dimension of this vector is defined as $d$, and is a hyperparameter of the system)
	\item $\pi_t$: The vector of fill probabilities corresponding to the orders represented by $u_t$, also of dimension $d$
	\item $c_t$: The corresponding vector of execution costs per unit order of $u_t$ at time $t$
	\item $s_{t+1}$: The scheduled position (as specified by TWAP, VWAP, etc)
	\item $\gamma$: A positive hyperparameter that weighs the schedule deviation penalty (for higher values of $\gamma$ the schedule is followed more closely)
	\item $\xi_t$: The per-share valuation or rollout cost per share for the expected residual quantity
\end{itemize}
In reference to the MPC equation \eqref{eq:approx_dp}, the state $x_t$ consists of $q_t$ together with the market state (the limit order book) at time $t$. The control is $u_t$ as defined above. The control constraint set $U_t(x_t)$ is defined by market and risk-related conditions at the current state (see the subsequent discussion in Section~\ref{subsubsec:constraints}). The probability vector $\pi_t$ encodes the uncertainty, and is appropriately estimated in our implementation (see Section~\ref{subsubsec:fill_probability}).

The schedule deviation at time $t+1$ is the random variable
$$\epsilon_{t+1} = q_{t+1} - s_{t+1}.$$
Its mean,
$$\hat{m}_t = q_t+\pi_t' u_t  - s_{t + 1},$$
appears in the schedule deviation penalty term in the minimization  Eq.~\eqref{eq:trading_problem}.

\section{Model Implementation}
\label{sec:model_implementation}

We now provide the implementation details of our MPC execution system. Consistent with the requirements listed in Section~\ref{subsec:our_approach}, the components of our implementation form a modular infrastructure that is data-driven, fast, and flexible. They work together to solve the optimization problem defined in \eqref{eq:trading_problem} at each time step $t$.

In Section~\ref{subsec:our_approach} we specified the requirement that our algorithm should fully exploit the multiple sources of liquidity available in a live trading environment. Each element of the control vector $u_t$ corresponds to the quantity allocated to one of $d$ pre-specified order templates. These templates are \emph{partially parameterized} in the sense that all non-quantity attributes of the order are fixed at optimization time, while the order size is determined by the optimizer.

In particular, at time $t$, we construct a vector of $d$ candidate orders $o_t$, where the $i$-th component $u_{i,t}$ of the control vector $u_t$ specifies the quantity allocated to the corresponding order template $o_{i,t}$. Each candidate order is defined as
\[
o_{i,t} = (p_i, \text{venue}_i, \text{type}_i, \ldots),
\]
where $p_i$ denotes the order price, the order duration is one time step, and the remaining fields specify the target venue (NASDAQ, NYSE, IEX, \emph{etc.}), order type (limit, market, immediate-or-cancel, \emph{etc.}), and any other required parameters apart from quantity.

\subsection{System Components}
\label{sec:system_components}

The MPC problem in \eqref{eq:trading_problem} is a simple program with quadratic cost and constraints, and inputs provided by a few key pieces of infrastructure. These are:

\begin{enumerate}

\item Scheduler function $s_{t+1} = F_s(t)$. This component performs a function similar to the `Macro-trader' of Li \textit{et al} \cite{li2022hierarchical}.

\item Candidate orders model $o_t = F_o(x_t,t)$. This component generates the set of candidate orders at time $t$.

\item Fill probability model $\pi_t = F_{\pi}(x_t, o_t).$ For example, a market order has fill probability = 1.0.

\item Fill covariance model $\Sigma_t = F_{\Sigma}(x_t, o_t)$; see Section \ref{subsubsec:fill_probability}.

\item Control constraints $U_t(x_t) = F_u(\Sigma_t, x_t)$. A series of constraints are maintained throughout the order management process, managed by this component.

\item Trading cost model $c_t = F_c(x_t, o_t)$. This models the trading cost per share of a candidate order.

\item Future cost per share, $\xi_t = F_\xi(x_t)$. This component estimates the per-share cost of the residual quantity to be traded, by following a simple base policy.

\end{enumerate}

\subsubsection{Scheduler}
\label{subsec:scheduler}

The scheduler component can choose $s_t$ statically or in response to market conditions. We have kept it static in our initial implementation, choosing to follow a VWAP profile that we pre-compute before the trading session. The VWAP implementation of $F_s$  is given by
$$F_s(t) = Q \frac{\hat{\nu}_t}{\hat{\nu}_T},$$
where  $\hat{\nu}_t$ is the cumulative volume forecast for time $t$. By contrast, the TWAP implementation is given by
$$F_s(t) = Q \frac{t}{T}.$$
Alternatively, we could use an Almgren-Chriss profile, or train some model to predict a suitable $s_t$ given the input state $x_t$, like Hendricks \& Wilcox \cite{hendricks2014reinforcement}. In such a case, $F_s$ becomes a function of $x_t$ and $t$.

\subsubsection{Constraints}
\label{subsubsec:constraints}
At each time step, the component $F_u(\Sigma_t, x_t)$ chooses a set of constraints $U_t(x_t)$ for the MPC optimization.

\begin{enumerate}
\item $u_t \geq  0$, all quantities are positive; we are not allowed to sell if the parent is a buy order, and vice versa.
\item $u_t \leq \kappa$, for an individual max order size $\kappa$ (say 50\% of $Q$).
\item $1' u_t + q_t \leq s_{t + 1} + \rho^{\text{upper}}_t$, an upper tube bound (which can start at say 20\% of the order quantity and decay to 0 at $t=T$).
\item $1' u^m_t + q_t \geq s_{t + 1} - \rho^{\text{lower}}_t$, a lower tube bound (which will force the optimization to choose higher cost, higher probability orders if it falls too far behind the schedule). The vector $u^m_t$ is the slice of $u_t$ that corresponds to (guaranteed fill) market orders.
\item $\hat{v}_t = u_t' \Sigma_t u_t \leq \beta$,  which constrains the uncertainty we wish to permit around our target schedule. Here $\beta$ is a scalar hyperparameter, and  $\Sigma_t$ is the fill covariance, i.e.,  the covariance of the order vector $u_t$; see the next section.
\end{enumerate}

\subsubsection{Fill Probability and Covariance Models}
\label{subsubsec:fill_probability}

The systems  $F_{\pi}$ and $F_{\Sigma}$ go hand in hand, but in principle can be modeled separately. There are plenty of ways to model fill probability, for example Maglaras \textit{et al} \cite{maglaras2022deep} who use a recurrent neural network (RNN) and the limit order book microstructure. 

For an illustration of the model of fill covariance, consider the case of two limit orders. Suppose there are two levels in the book, one closer to the mid (level 1) and one deeper in the book (level 2). Let $z_1$ and $z_2$ be non-independent Bernoulli random variables corresponding to whether level 1 and level 2 are (fully) filled respectively, with probabilities
$$P\{z_1=1\} = \pi_1, \qquad P\{z_2=1\} = \pi_2.$$
The covariance between them is
$$\Sigma_{12} = \operatorname{Cov}(z_1, z_2) = E\{z_1 z_2\} - E\{z_1\}\cdot E\{z_2\},$$
where $\operatorname{Cov}(z_1, z_2)$ represents the covariance between the random variables $z_1$ and $z_2$. The matrix of joint outcome probabilities is
\[
\begin{array}{c|c|c}
& z_1=0 & z_1=1 \\
\hline
z_2=0 & 1 - \pi_1 & \pi_1 - \pi_2 \\
\hline
z_2=1 & 0 & \pi_2 \\
\end{array}
\]
Note that the entries sum to $1$. The asymmetry is due to the fact that it is impossible to fill the deeper level without also filling the shallow one. From this matrix,
$$E\{z_1 z_2\} = 0 \cdot (1 - \pi_1) + 0 \cdot 0 + 0 \cdot (\pi_1 - \pi_2) + 1 \cdot \pi_2 = \pi_2,$$
so that
$$\operatorname{Cov}(z_1, z_2)=E\{z_1 z_2\} - E\{z_1\}\cdot E\{z_2\} = \pi_2 - \pi_1 \pi_2.$$
Generalizing, for arbitrary levels $i$ and $j$, the joint probability of both being filled equals the probability of filling the deeper level, i.e.
$$E\{z_i z_j\}  = \min\{\pi_i, \pi_j\}.$$
Therefore a model of the fill covariance matrix for these orders is
$$\Sigma_{ij} = \min\{\pi_i, \pi_j\} - \pi_i \pi_j,$$
which we can use to construct $F_{\Sigma}$. Note that in practice orders can be partially filled; this is a simple model used to bootstrap our system. For more complex order types (and venues) we can measure the fill covariance empirically.

A more complex fill probability model involves conditioning on a \emph{fast cancel} mechanism. Rather than modeling the unconditional fill probability $\pi_i = P\{z_i = 1\}$, we instead model the conditional probability
$$\pi_i^{c} = P\{z_i = 1\mid \text{not cancelled}\},$$
where cancellation is triggered by a separate module that monitors adverse market conditions in real time.\footnote{In practice, this type of system usually runs on an ultra-low latency field-programmable gate array (FPGA).}

The fast-cancel module needs to operate at very low latency and withdraws resting limit orders when microstructural signals indicate imminent adverse selection. A simple signal is to trigger when the order book imbalance shifts sharply. This creates a conditional fill distribution that is substantially more favorable (with respect to adverse selection) than the unconditional one: fills that would have occurred just before a price move we would benefit from are systematically avoided.

From a modeling perspective, this decomposition is advantageous because the conditional fill probability $\pi_i^{c}$ can be learned from historical data where the fast-cancel logic was active. The resulting model captures the effective fill dynamics experienced by the trading system in production. This approach separates concerns: the fill probability model $F_\pi$ estimates execution likelihood given that orders remain active, while the fast-cancel module independently manages adverse selection risk. Both components can be trained and improved in isolation, consistent with the modularity requirements outlined in Section~\ref{subsec:our_approach}. All other components remain unchanged.

\subsubsection{Trading Cost per Share}
\label{subsubsec:order_pricing}
Next, we discuss the trading cost per share. The simplest approach is to represent this cost as a vector of components, with the component for candidate limit order $o_i$ being  the deviation of the  order from the market price, normalized by spread:
$$\phi\, \frac{p_i - p^m_t}{\delta_t} $$
where $\phi$ is the side multiplier (=1 for buys, -1 for sells).
Here $p_i$ is the price attached to $o_i$. It is set to $p^m_t + 0.5 \cdot \delta_t$ for market orders where we cross the spread (and don't attach a limit price), or simply to $p^m_t$ for a mid-IOC (immediate-or-cancel order).

This form of $F_c$ measures the price paid (in units of spread) relative to the mid. Alternative cost functions exhibit similar properties. For example, consider a cost defined as the difference between the trade price and the $t+1$ mid price (often referred to in the industry as a `markout'). Both specifications yield an increasing cost as a function of fill probability. This pattern is driven by market microstructure. Heuristically, consider two cases: (A) we are filled at the front of a long, stable queue; (B) we are filled at the end of the queue as a level collapses. In case (A) we collect half the spread relative to the mid at the end of the interval, for case (B) we pay it. Lower probability orders, deeper in the book, are more likely to be filled at stable levels. Equivalently, very passive orders tend to have lower market impact than aggressive ones. Given the similarity in properties, we adopt the simpler cost function for speed and interpretability. There are other sensible order pricing methods; we could train a neural network to predict an interval VWAP slippage for the candidate order $o_{i, t}$, or map the cost function to a traditional impact model, such as after Cont \textit{et al} \cite{cont2014price}. We consider this an open research question.

\subsubsection{Future Cost per Share}
\label{subsubsec:rollout_cost}
The future cost approximation is set equal to the cost of following a simple base policy. A scalar $\xi_t$ encodes the expected (per-share) cost of executing the remaining shares under this policy, expressed in units of spreads (consistent units with $F_c$). For simplicity, we consider a base policy that submits market orders for the residual quantity, yielding $\xi_t = 0.5$, i.e., half a spread. This mechanism accommodates price forecasts: if we expect prices to move against us, we can increment $\xi_t$ accordingly. This is discussed in Section \ref{subsec:base_policy}.

\subsection{Optimization}
\label{subsec:optimization}

Bringing all of this together, we substitute terms into \eqref{eq:trading_problem} to obtain the final quadratic programming problem solved at each time step $t$. Our implementation of this problem uses the fast second order conic solver Clarabel \cite{Clarabel_2024}, and takes about 1 millisecond\footnote{Experiments were conducted on a server equipped with two AMD EPYC 7R13 processors.} for an action space of $d=11$, which is consistent with requirement 2 of Section~\ref{subsec:our_approach}.

\section{Experiments}
\label{sec:experiments}

In this section we discuss our algorithm's performance in a simulated environment.

\subsection{Simulation Environment}
\label{subsec:simulation_environment}

We trade 1200 instruments per day on a simulated NASDAQ. For each instrument on each trading day for six months (2025-01-02 to 2025-07-02) we manage a \$10K parent order over the full session, alternating buying and selling each day. This corresponds to $\approx$ 170,000 parent orders. For each instrument we maintain a full order book, built using L3 ITCH message data. Additionally, we simulate a conservative 10 ms latency between order submission and interaction with the book. In our simulation environment we can submit both market and limit orders. Market orders remove liquidity upon arrival, limit orders join or create a price level queue. We simulate limit order fills when the order behind us in the queue gets filled. If the filled quantity is less than our order quantity, we partially fill and leave the residual resting in the queue.

\subsection{Metrics}
\label{subsec:metrics}

We evaluate execution performance using three complementary price-based metrics, designed to isolate different aspects of execution quality. All metrics are expressed in basis points (bps) and normalized so that positive values correspond to worse execution outcomes for both buy and sell orders.
Let:
\begin{itemize}
  \item $p_0$ denotes the \emph{arrival price}, defined as the mid-price at the time the parent order is received;
  \item $p_{\text{vwap}}$ denotes the \emph{market VWAP}, defined as the volume-weighted average traded price over the lifetime of the order;
  \item $p_{\text{fwap}}$ denotes the \emph{fill-weighted average price (FWAP)}, defined as the quantity-weighted average price of all executed trades generated by the algorithm across one parent order;
  \item $p_{\text{swap}}$ denotes the \emph{schedule-weighted average price (SWAP)}, defined as the hypothetical FWAP that would be obtained if the prescribed execution schedule were followed exactly and all scheduled quantities were executed at the current mid-price at each decision time.
\end{itemize}
Using the side multiplier $\phi$ (=1 for buys, -1 for sells) these define the following metrics:

\begin{itemize}

\item Arrival slippage (bps):
$$z_{\text{arrival}} = 10,000\, \frac{p_{\text{fwap}} - p_0}{p_0} \phi$$
Measures how far our realized execution price drifted from the price when the order arrived.

\item VWAP slippage (bps):
$$z_{\text{vwap}} = 10,000\, \frac{p_{\text{fwap}} - p_{\text{vwap}}}{p_{\text{vwap}}} \phi$$
Compares our execution price against the market average price over the same window.
\item Schedule shortfall (bps):
$$z_{\text{schedule}} = 10,000\, \frac{p_{\text{fwap}} - p_{\text{swap}}}{p_{\text{swap}}} \phi$$
Measures how much worse (or better) our algorithm performed relative to its own intended schedule.
\end{itemize}

\subsection{Performance Across Schedule Types}
\label{subsec:performance_schedules}

TWAP, VWAP, and Almgren-Chriss trading schedules are defined as follows:

\begin{itemize}

\item TWAP: We trade linearly in time:
$$s_t = Q\frac{t}{T}$$

\item VWAP: We trade along the schedule of an offline volume profile forecast (that uses information up to the previous trading day):
$$s_t = Q \frac{\hat{\nu}_t}{\hat{\nu}_T}$$

\item Almgren-Chriss: We trade with a static Almgren-Chriss (after \cite{almgren2000optimal}) profile, wrapping the impact terms into a single shared parameter $\psi$ that we illustratively set to 5 basis points:
$$s_t = Q \left[1 - \frac{\sinh(\psi(T - t))}{\sinh(\psi T)}\right]$$

\end{itemize}
Unless otherwise noted, all experiments in Section~\ref{subsec:performance_schedules} use the same optimization, solver, hyperparameters, and candidate order set; only the scheduler $F_s$ (hence $s_t$) differs across TWAP, VWAP, and Almgren-Chriss. This shared parameterization is shown in Table \ref{tab:sim_params}.

\begin{table}[ht]
\centering
\begin{tabular}{l l}
	\toprule
	Parameter & Value \\
	\midrule
	$\text{Interval}$ & $\text{5 minutes}$ \\
	$\xi_t$ & $0.5$ \\
	$\rho_t^{\text{upper}}, \rho_t^{\text{lower}}$ & 15\% \\
	$\beta$ & 5 \\
	$\gamma$ & 1 \\
	$d$ & {11} \\
	$o_{t, 0}$  & Market order \\
	$o_{t, i}$  & Increasingly passive limit orders, \\
    &$i=1,\dots,10$ \\
	$\pi_{t, 0}$ & Market order fill probability,  1.0 \\
	$\pi_{t, i}$ & Linearly decreasing from $0.9$ to $0.1$,\\ &$i=1,\dots,10$ \\
	\bottomrule
\end{tabular}
\caption{Simulation baseline parameters}
\label{tab:sim_params}
\end{table}

\subsubsection{Cost of Execution}
\label{subsubsec:cost_of_execution}

Slippage measurements are made for the three schedules to illustrate the flexibility of our formulation. We do not wish to compare performance across profile types. These structures are chosen in a live setting to minimize market impact, which we are not simulating. Our choice of a static 5 bps for the Almgren-Chriss parameter, for example, is arbitrary and should be refined per-instrument.

For each candidate profile we run two simulations, one using the MPC optimization procedure detailed in Section~\ref{sec:problem_formulation}, and one that crosses the spread at each optimization step for the scheduled quantity (labeled `crossing').

Slippage metrics from the simulations are summarized in Table \ref{tab:performance}, but we draw more attention to Table \ref{tab:relative_performance}, which compares performance for each scheduling type to its spread-crossing baseline.

For clarity and consistency across tables and figures, we use the following naming convention for all execution policies considered:
\begin{itemize}
  \item TWAP/Schedule, VWAP/Schedule, AC/Schedule: the schedule being followed, independent of any policy.
  \item TWAP/MPC, VWAP/MPC, AC/MPC: the proposed MPC execution method, using the corresponding schedule.
  \item TWAP/Crossing, VWAP/Crossing, AC/Crossing: the spread-crossing baseline that executes the scheduled quantity at each decision time by crossing the spread.
  \item TWAP/Oracle, VWAP/Oracle, AC/Oracle: the MPC method with an oracle base policy that uses future price information (e.g., close price) to compute the future cost.
\end{itemize}

\begin{table}[ht]
\centering
\begin{tabular}{lrrr}
	\toprule
	Cost / bps & $z_{\text{arrival}}$ & $z_{\text{schedule}}$ & $z_{\text{vwap}}$ \\
	\midrule
	{VWAP/MPC}        & 18.10 &  4.53 &  4.36 \\
	{VWAP/Crossing} & 19.59 &  6.75 &  6.12 \\
	{TWAP/MPC}   & 16.98 &  4.71 &  5.49 \\
	{TWAP/Crossing} & 19.15 &  6.75 &  6.83 \\
	{AC/MPC}          & 20.99 &  9.03 & 12.46 \\
	{AC/Crossing}  & 21.70 & 13.21 & 17.22 \\
	\bottomrule
\end{tabular}
\caption{Performance metrics (in basis points) across strategies. 	Positive values indicate worse execution for both buys and sells.}

\label{tab:performance}
\end{table}

\begin{table}[ht]
\centering
\begin{tabular}{lrrr}
	\toprule
	Improvement / \% & $\Delta z_{\text{arrival}}$ & $\Delta z_{\text{schedule}}$ & $\Delta z_{\text{vwap}}$ \\
	\midrule
	VWAP/MPC &  8.23 & 48.85 & 40.37 \\
	TWAP/MPC & 12.77 & 43.14 & 24.55 \\
	AC/MPC   &  3.37 & 46.34 & 38.20 \\
	\bottomrule
\end{tabular}
\caption{Performance improvements of the MPC policies over the spread-crossing baseline for each profile type.}
\label{tab:relative_performance}
\end{table}

The results of Tables \ref{tab:performance} and \ref{tab:relative_performance} clearly demonstrate that the MPC algorithm provides a significant performance boost, including greater than 40 \% improvement in the cost of following each candidate schedule. Arrival and VWAP improvements are more varied; these are more of a function of the profiles themselves relative to actual market moves.

\subsubsection{Accuracy of Schedule Following}
\label{subsec:accuracy_schedule}

\begin{figure*}[p]
	\centering
	\includegraphics[width=\textwidth]{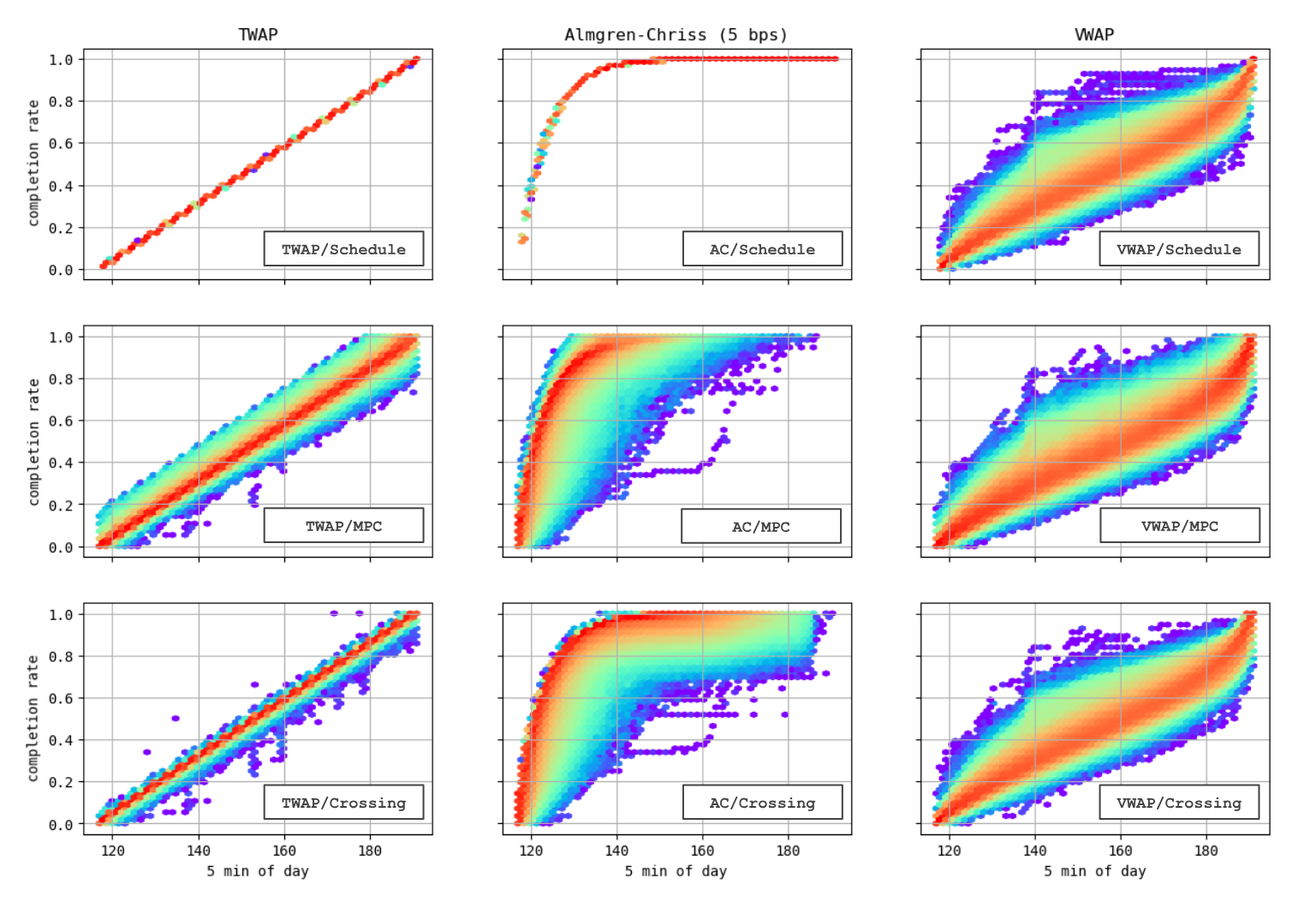}
	\caption{Schedule following across three candidate profile types. Using MPC vs simply crossing the spread. Red corresponds to higher density.}
	\label{fig:schedule_following}
\end{figure*}

\begin{figure*}[p]
	\centering
	\includegraphics[width=\textwidth]{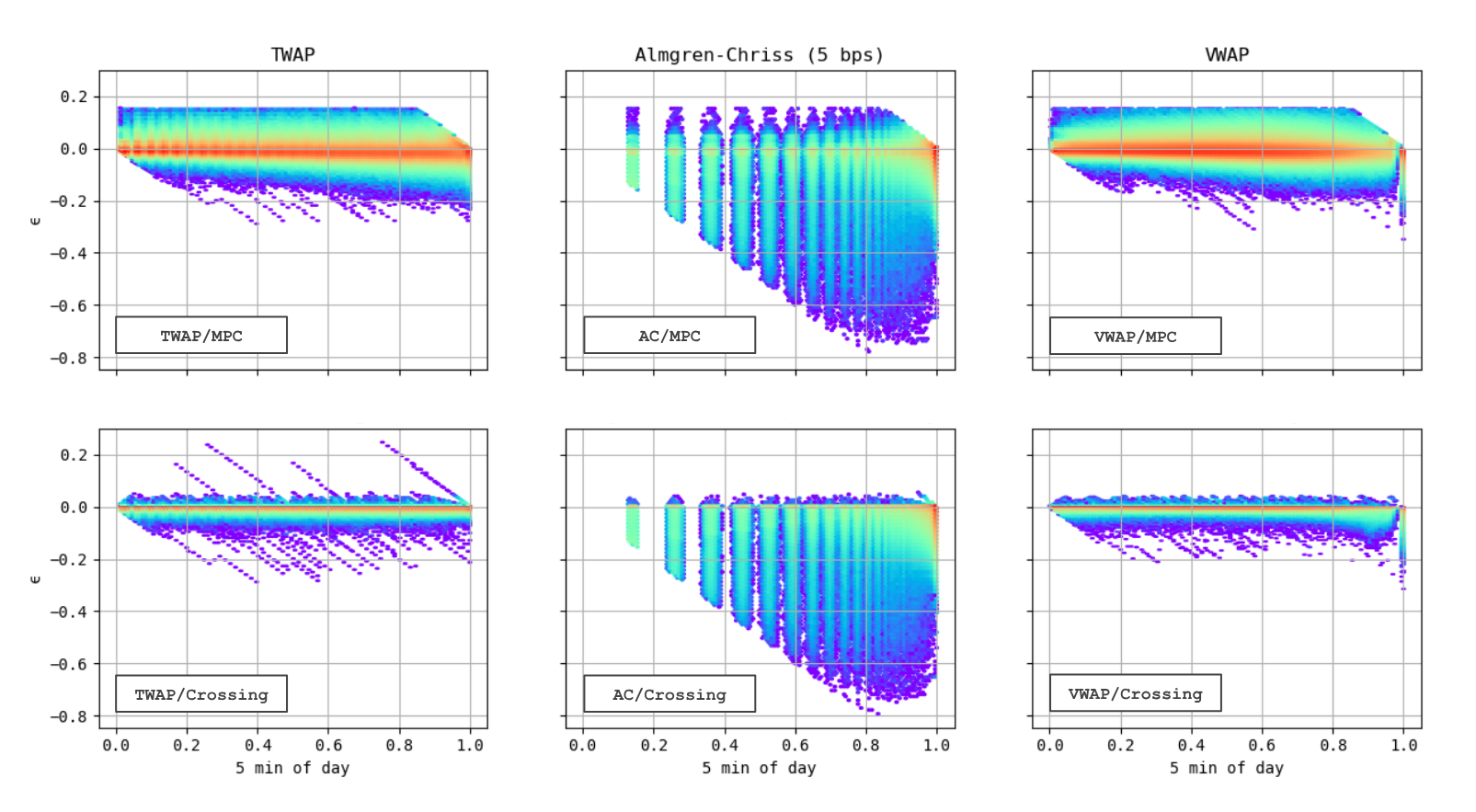}
	\caption{Evolution of schedule deviation across the trading day.}
	\label{fig:schedule_deviation}
\end{figure*}

\begin{figure*}[p]
	\centering
	\includegraphics[width=\textwidth]{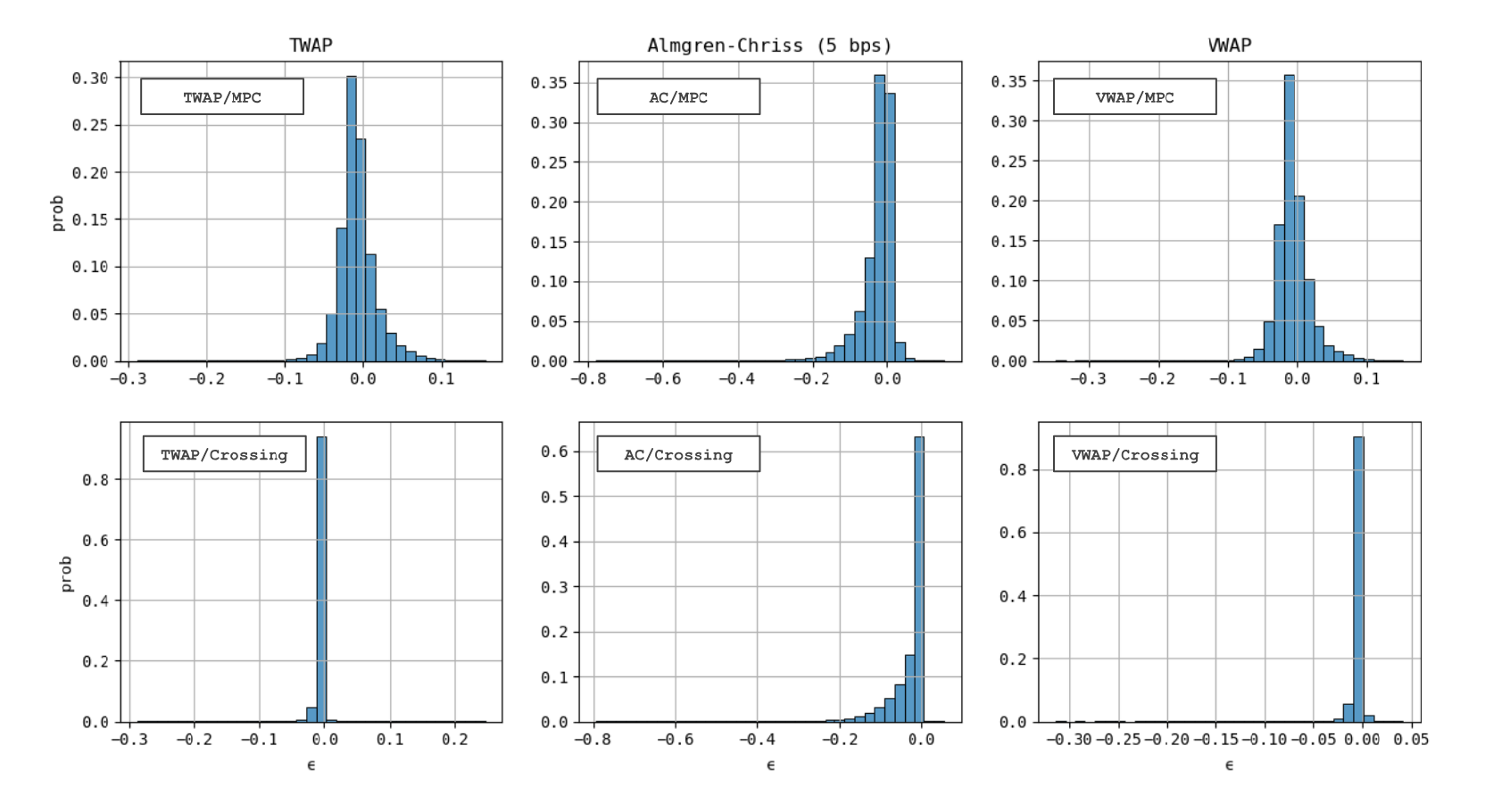}
	\caption{Histograms of schedule deviation.}
	\label{fig:schedule_deviation_histogram}
\end{figure*}

Figure \ref{fig:schedule_following} shows average intraday completion rate densities. The optimization maintains a stable tube around the schedule, most evidently for TWAP policies. This is clearer in Figure \ref{fig:schedule_deviation} for the Almgren-Chriss and VWAP profiles. Summary statistics for these deviations are presented in Table \ref{tab:exec-stats}, with corresponding histograms shown in Figure \ref{fig:schedule_deviation_histogram}.

\begin{table}[h]
\centering
\small
\begin{tabular}{lrrr}
	\toprule
	$\epsilon_t$ / \% & Mean & Std & Median \\
	\midrule
	{TWAP/MPC}                 & -0.765 &  2.471 & -1.062 \\
	{TWAP/Crossing }     & -0.271 &  0.534 & -0.030 \\
	{AC/MPC}       & -2.849 &  4.807 & -1.459 \\
	{AC/Crossing} & -2.756 &  4.827 & -0.712 \\
	{VWAP/MPC}                 & -0.574 &  2.450 & -0.912 \\
	{VWAP/Crossing}      & -0.231 &  0.545 & -0.024 \\
	\bottomrule
\end{tabular}
\caption{Schedule deviation summary statistics.}
\label{tab:exec-stats}
\end{table}

On average, all simulations slightly lag the prescribed schedule. For the crossing simulations, this is attributed due to our simulated latency. Specifically, price movements in the trading direction between optimization and order submission can result in some market orders remaining unfilled, since such orders are simulated as limit orders placed at the far touch. In the MPC simulations, the lag is instead attributed to imperfect calibration of fill probabilities: although the optimizer targets schedule adherence in expectation, realized executions tend to underfill with this fill probability model.

The Almgren-Chriss profiles accelerate aggressively, resulting in a substantially higher concentration of market orders, particularly at the beginning of the trading session. This behavior is reflected in larger negative schedule deviations and, more generally, inferior slippage performance (see Table~\ref{tab:performance}).

\begin{figure*}[p]
	\centering
	\includegraphics[width=\textwidth]{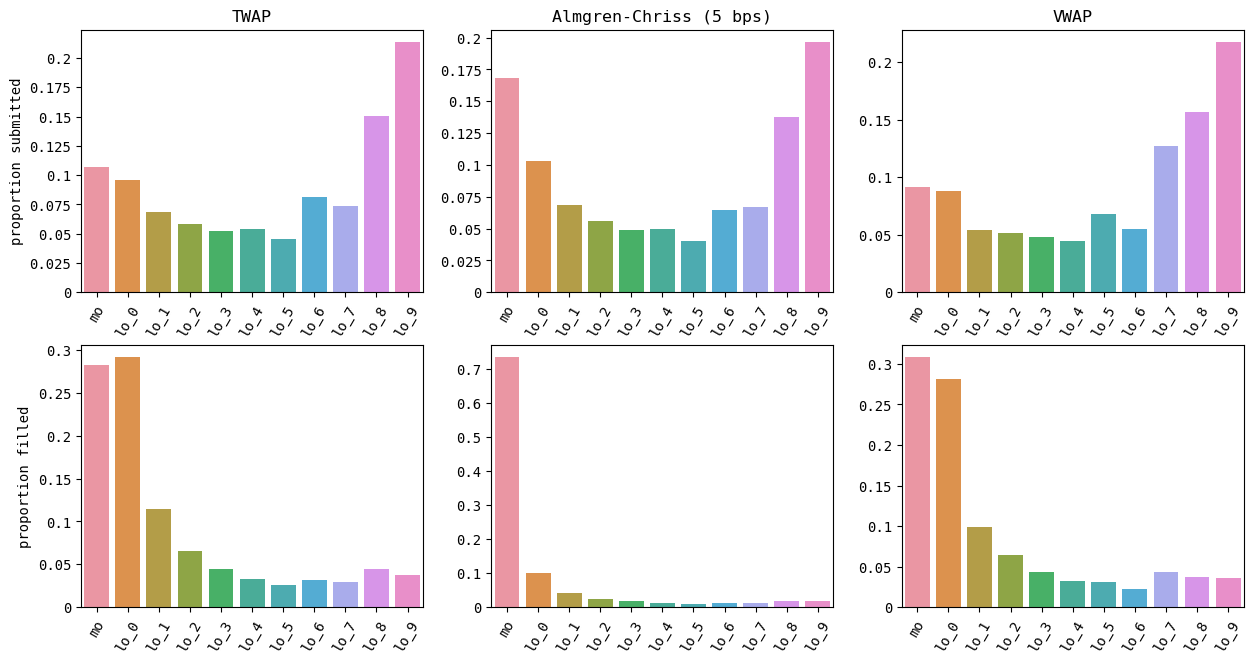}
	\caption{Submitted and filled price levels}
	\label{fig:levels}
\end{figure*}

Figure \ref{fig:levels} shows the distribution of quantities submitted (values of the chosen action vector $u_t$) for the MPC simulations. Significantly higher density of market orders for Almgren-Chriss profiles is represented here, both for submitted and filled. The TWAP and VWAP profiles exhibit similar behavior, allocating as much quantity as feasible to low-probability, high-payoff orders, as previously discussed in Section \ref{subsubsec:order_pricing}.

\subsection{Hyperparameter Selection}
\label{sec:expt_hyperparameter}

The hyperparameters $\gamma$ and $\beta$ play an important role in controlling the optimization. In particular, $\gamma$ controls how strongly $|\hat{m}_t|^2$ is penalized. This is a soft constraint; the optimizer is free to target positions above or below the schedule (subject to other constraints) depending on the value of $\xi_t$. If it is relatively cheap to execute more shares at the current stage, then it may be desirable to have positive expected schedule deviation $E\{\epsilon_{t+1}\} > 0$ (and the opposite for a relatively expensive current stage).

In contrast, $\beta$ controls the amount of `risk' the optimizer can take. Higher $\beta$ encourages greater concentration of order quantity on the lower probability but better payoff price levels.

\begin{figure*}[p]
	\centering
	\includegraphics[width=\textwidth]{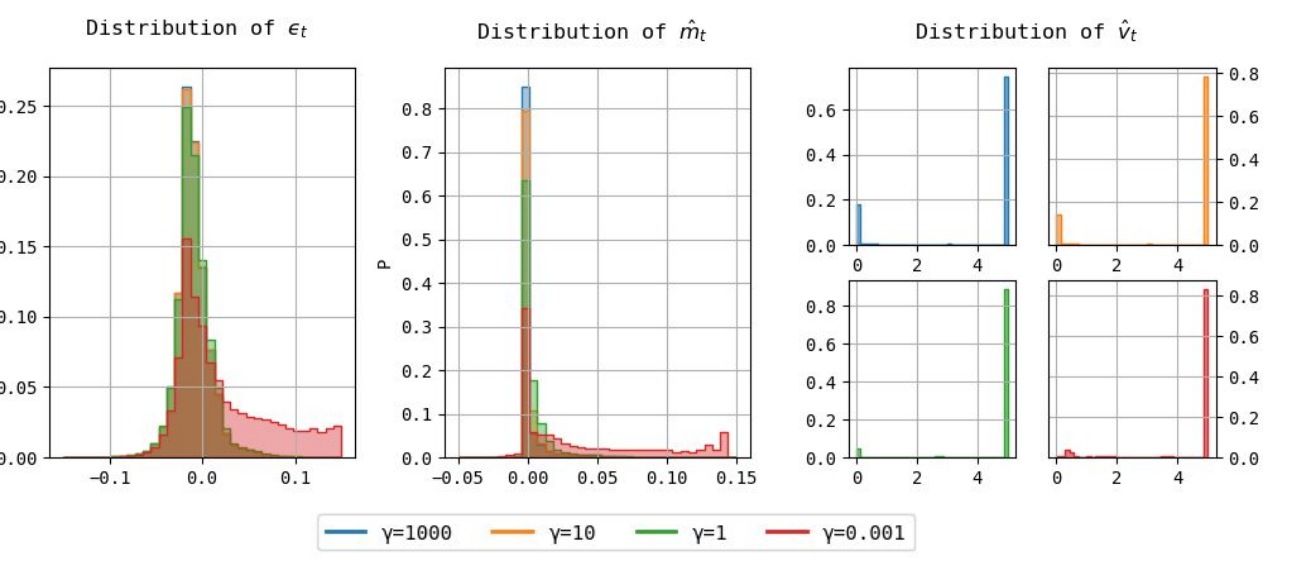}
	\caption{Histograms of $\epsilon_t$ and targeted moments as a function of $\gamma$.}
	\label{fig:gamma_eps_moments}
\end{figure*}

\begin{figure*}[t]
	\centering
	\includegraphics[width=\columnwidth]{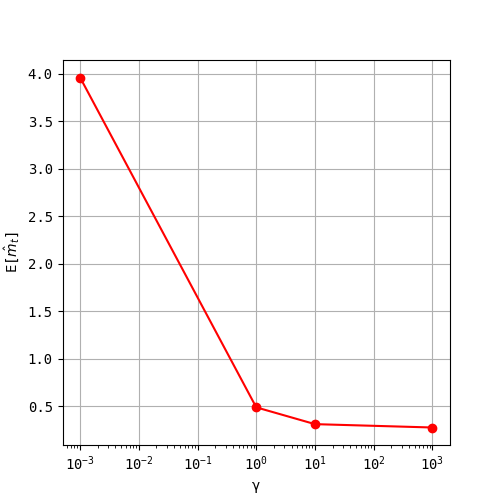}
	\caption{$E\{\hat{m}_t\}$ vs target $\gamma$.}
	\label{fig:gamma_m_hat}
\end{figure*}

Using the same base parameterization as described in Table \ref{tab:sim_params}, and a TWAP profile, Figure \ref{fig:gamma_eps_moments} shows the distributions of $\epsilon_t$, $\hat{m}_t$ , and $\hat{v}_t$ across simulations as we vary $\gamma$. Under the spread-crossing base policy, $\hat{m}_t$ is almost always positive, as the cost of rollout is higher than any action at the current stage. As $\gamma$ is increased the density of $\hat{m}_t$ increasingly clusters at zero. This is shown clearly in Figure \ref{fig:gamma_m_hat}.

\begin{figure*}[p]
	\centering
	\includegraphics[width=\textwidth]{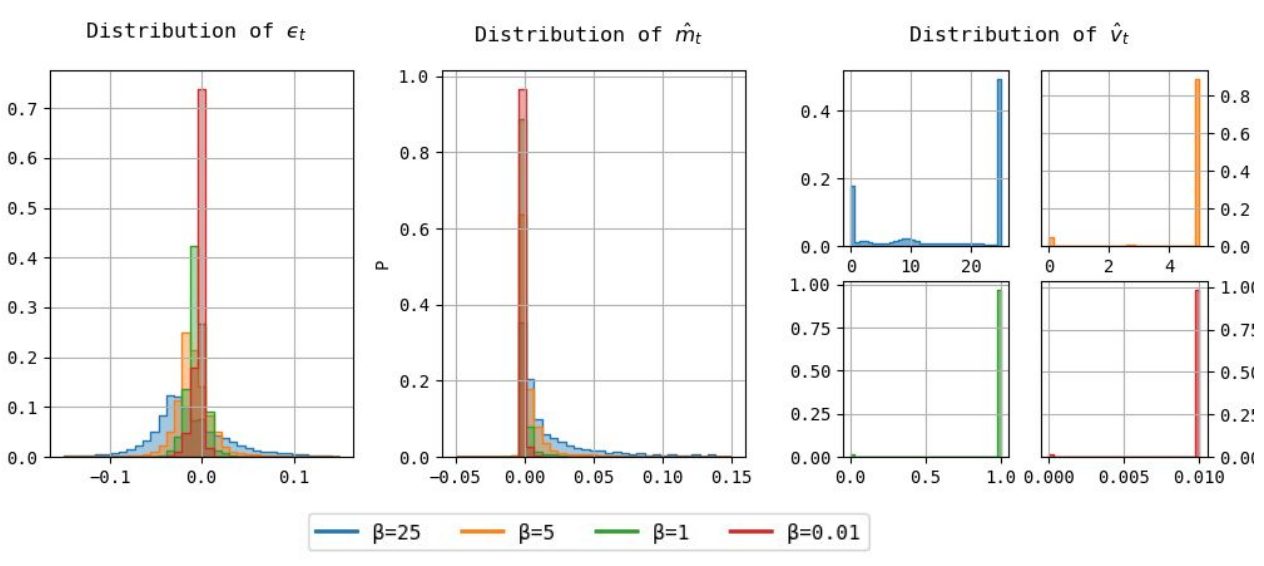}
	\caption{Histograms of $\epsilon_t$ and targeted moments as a function of $\beta$.}
	\label{fig:beta_eps_moments}
\end{figure*}

\begin{figure*}[t]
	\centering
	\includegraphics[width=\columnwidth]{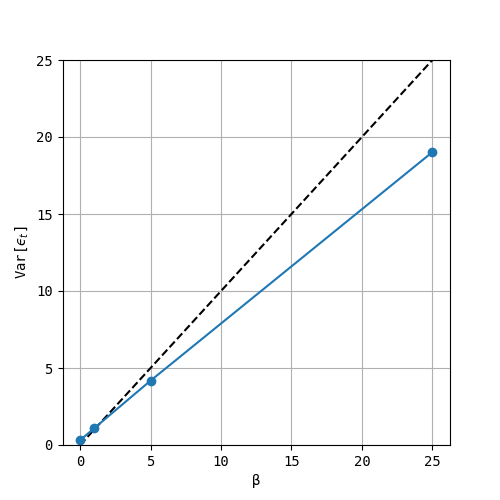}
	\caption{Var$\{\epsilon_t\}$ vs target $\beta$. The dotted line shows perfect calibration.}
	\label{fig:beta_calib}
\end{figure*}

Figure \ref{fig:beta_eps_moments} shows the schedule deviation $\epsilon_t$ and its target moments across simulations as we now vary $\beta$. The intention here is to control the variance of $\epsilon_t$, denoted by $\text{Var}\{\epsilon_t\}$, and it is clear that as $\beta$ increases the distribution of $\epsilon_t$ widens. To verify the calibration of our control, we plot realised $\text{Var}\{\epsilon_t\}$ as a function of $\beta$ across the same simulations in Figure \ref{fig:beta_calib}. At higher values of $\beta$ the amount of risk we can practically take seems to be limited by the outer tube ($\rho_{upper}, \rho_{lower}$), but in these simulations we observe good calibration. To confirm that the risk we are taking is worth it, we plot improvement in slippage metrics (over the spread crossing baseline, as in Table \ref{tab:relative_performance}) as a function of $\beta$. Taking risk pays off, and performance monotonically improves across all metrics as $\beta$ increases. This can be attributed to an increased density of (filled) low probability and high payoff orders.

\subsection{Base Policy Design}
\label{subsec:base_policy}

So far we have used a spread-crossing base policy for rollout and defined $\xi_t = 0.5$ (half a spread). This is a simple but quite pessimistic choice. The role of the base policy is to provide a mechanism to approximate $J^*_t(x_t)$. In practical trading, there may be a short term price forecast that we wish to incorporate. If we predict that the price will increase over the next 5 minutes (and we're buying), we can incorporate this information into the rollout cost $\xi_t$. The base policy becomes ``cross the spread at our forecasted price level." This has the effect of increasing the cost of the residual quantity relative to executing these shares now, encouraging the optimization to exceed the schedule, which is desirable behavior.

To demonstrate this effect, we keep other simulation and optimization parameters constant, then measure performance of an `oracle' base policy by setting 
$$\xi_t = \phi\,{p_{\text{close}}-p^m_t\over \delta_t},$$ where $\phi$ is the side multiplier and $p_{\text{close}}$ is defined as the daily closing auction price (which happens at the end of the trading session, after our order completes).

\begin{table}[ht]
\centering
\begin{tabular}{lrrr}
	\toprule
	Cost / bps & $z_{\text{arrival}}$ & $z_{\text{schedule}}$ & $z_{\text{vwap}}$ \\
	\midrule
	{TWAP/MPC}        & 16.98 &  4.71 &  5.49 \\
	{TWAP/Crossing} & 19.15 &  6.75 &  6.83 \\
	{TWAP/MPC-Oracle} & 7.298 &  -5.62 &  -5.42 \\
	\bottomrule
\end{tabular}
\caption{Performance comparisons with the oracle policy as the base policy. Here we use `MPC-Oracle' to denote the MPC policy with the oracle base policy.}
\label{tab:oracle_performance}
\end{table}

\begin{table}[ht]
\centering
\begin{tabular}{lrrr}
	\toprule
	Improvement / \% & $\Delta z_{\text{arrival}}$ & $\Delta z_{\text{schedule}}$ & $\Delta z_{\text{vwap}}$ \\
	\midrule
	{TWAP/MPC} & 12.77 & 43.14 & 24.55\\
	{TWAP/MPC-Oracle} & 162.43 & 220.06 & 226.06 \\
	\bottomrule
\end{tabular}
\caption{Performance improvement over the spread-crossing baseline for the MPC and the MPC-Oracle policies.}
\label{tab:oracle_relative_performance}
\end{table}
Tables \ref{tab:oracle_performance} and \ref{tab:oracle_relative_performance} demonstrate the significant improvement achieved when we provide the optimization with future price information. Though obviously this approach is not realistic ($p_{\text{close}}$ is not known), our results demonstrate that inclusion of accurate price predictions into the base policy can yield significant performance improvements.

\section{Discussion}
\label{sec:discussion}

Our MPC algorithmic framework for trade execution balances schedule following with controlled risk taking. It is modular, fast, and agnostic to the chosen execution schedule. The experimental results we have presented demonstrate significant performance improvements for a static strategy, trading \$10K alternating buy and sell parent orders, across three schedule types (TWAP, VWAP, and Almgren-Chriss).

As described in Section~\ref{sec:system_components}, in a live trading environment we can exploit a substantially richer action space than simple limit and market orders. In addition to direct order placement, a wide range of execution broker algorithms is available (see, e.g.,~\cite{morganstanley2024-us-cash-equity-faq, jpmorgan2019-fx-ml, ubs2019-fx-algos}), as well as multiple trading venues. In our framework, each such order configuration corresponds to an element of the vector $o_t$ at optimization time. We can therefore assign a cost to each candidate order type that more accurately reflects its realized cost as a function of the current market state $x_t$. For example, passive limit orders typically offer more favorable payoffs when filled, while VWAP-following broker algorithms tend to perform better when prices are trending away from the trader. These effects can be learned from data, allowing the use of an offline (or indeed online) trained model for $F_c$ (as a function of broker, venue, state, etc) rather than the simplified aggressiveness-based cost specification employed in our simulations.

The choices of the hyperparameters $\beta$ and $\gamma$ are important, as discussed in Section~\ref{sec:expt_hyperparameter}, and we observe effective control of the mean and variance of schedule deviation. However, we note that these constraints are only required due to our limited lookahead: they constrain the action search space to regions we believe will perform well over the full horizon. A better approximation to the expected future cost, possibly through an improved transition function (in the literature often referred to as a `world model'), would enable further lookahead, allowing the optimizer to properly evaluate the consequences of actions. In turn, this reduces the need to constrain the search.

Our formulation also admits an extension in which an \emph{outer-loop} controller selects the optimization hyperparameter tuple $(\beta, \gamma, \rho_t^{\text{upper}}, \rho_t^{\text{lower}}, \xi_t)$ as an action, based on the same observed market state $x_t$. Such a mechanism would allow the system to adapt its risk profile dynamically, taking on greater risk in more benign market conditions. Additionally, we note that the framework is easily extended to accommodate orders with durations spanning multiple time steps. We note these possibilities here and leave their implementation and empirical evaluation for future work.

\section{Conclusions}
\label{sec:conclusions}

We have presented an MPC-based framework of schedule-informed parent order execution. It is free of any market dynamics modelling, scalable, and modular. Using NASDAQ L3 simulations we have shown:

\begin{itemize}
\item Explicit control of expected schedule deviation and its uncertainty, governed by two hyperparameters, $\beta$ and $\gamma$,  and demonstrated it across three schedule types: TWAP, VWAP and Almgren-Chriss.

\item Significant performance improvement across three slippage metrics (arrival slippage, VWAP slippage, schedule shortfall) in comparison to a spread-crossing baseline.

\item Even greater performance gains are observed when the future cost is computed  through an \emph{oracle} base policy that uses estimated future price information. In this setting, the algorithm effectively balances current-stage execution costs and expected schedule deviation against anticipated closing prices, resulting in substantially more efficient trading. While such oracle information is not available in practice, our results suggest that incorporating short-horizon price forecasts (possibly generated by a neural network) into the rollout component may yield significant benefits.

\end{itemize}

\bibliography{references}

\clearpage
\onecolumn
	
\appendix

\section{Notation Reference}
\label{sec:append_ref}

\renewcommand{\arraystretch}{1.1} 

\begin{table}[h!]
\centering
\caption{Notation Reference}
\label{tab:notation_reference}
\small
\begin{tabular}{ll}
	\toprule
	\textbf{Symbol} & \textbf{Description} \\
	\midrule
	$T$ & Duration of the parent order (number of time periods) \\
	$t$ & Time index, $0 \leq t \leq T$ \\
	$x_t$ & State at time $t$ (includes $q_t$ and market state) \\
	$u_t$ & Control vector of order quantities at time $t$ \\
	$w_t$ & Random quantity modeling uncertainty \\
	$f_t$ & System dynamics function \\
	$g_t$ & Stage cost function \\
	$J^*_t(x_t)$ & Optimal cost-to-go from state $x_t$ \\
	$\tilde{J}_{t+1}$ & Approximate cost-to-go (rollout approximation) \\
	$U_t(x_t)$ & Control constraint set at time $t$ \\
	\midrule
	$Q$ & Quantity of the parent order \\
	$q_t$ & Executed position at time $t$ \\
	$s_t$ & Scheduled position at time $t$ \\
	$d$ & Dimensionality of action space (number of candidate orders) \\
	$o_{i,t}$ & Candidate order $i$ at time $t$ \\
    $p_i$ & Limit price of candidate order $o_i$ \\
	$\pi_t$ & Fill probability vector \\
	$\Sigma_t$ & Fill covariance matrix \\
	$c_t$ & Trading cost vector per share (in units of spreads) \\
	$\kappa$ & Maximum individual order size \\
	$\delta_t$ & Bid-ask spread at time $t$ \\
	$p^m_t$ & Mid price at time $t$ \\
	$p_{\text{close}}$ & Close (future) price \\
	\midrule
	$\gamma$ & Schedule deviation penalty hyperparameter \\
	$\beta$ & Variance constraint hyperparameter \\
	$\xi_t$ & Rollout cost per share (in units of spreads) at time $t$ \\
	$\psi$ & Almgren-Chriss impact parameter \\
	$\rho^{\text{upper}}_t, \rho^{\text{lower}}_t$ & Upper and lower tube bounds \\
	$\hat{\nu}_t$ & Cumulative volume forecast for time $t$ \\
	\midrule
	$\epsilon_{t+1}$ & Schedule deviation at time $t+1$ ($q_{t+1} - s_{t+1}$) \\
	$\hat{m}_t$ & Expected schedule deviation, $E\{\epsilon_{t+1}\}$ \\
	$\hat{v}_t$ & Variance of schedule deviation, $\text{Var}\{\epsilon_{t+1}\}$ \\
	\midrule
	$p_0$ & Arrival price (mid-price when order received) \\
	$p_{\mathrm{fwap}}$ & Fill-weighted average price \\
	$p_{\mathrm{vwap}}$ & Market volume-weighted average price \\
	$p_{\mathrm{swap}}$ & Schedule-weighted average price \\
	$\phi$ & Side multiplier ($+1$ buy, $-1$ sell) \\
	$z_{\text{arrival}}$ & Arrival slippage (bps) \\
	$z_{\text{vwap}}$ & VWAP slippage (bps) \\
	$z_{\text{schedule}}$ & Schedule shortfall (bps) \\
	\midrule
	$F_s$ & Scheduler function \\
    $F_o$ & Candidate order generator \\
	$F_u$ & Constraint controller \\
	$F_\pi$ & Fill probability model \\
	$F_\Sigma$ & Fill covariance model \\
	$F_c$ & Order trading cost model \\
	$F_\xi$ & Rollout cost model \\
	\bottomrule
\end{tabular}
\end{table}

\clearpage
\twocolumn

\begin{figure*}[t]
	\centering
	\includegraphics[width=\columnwidth]{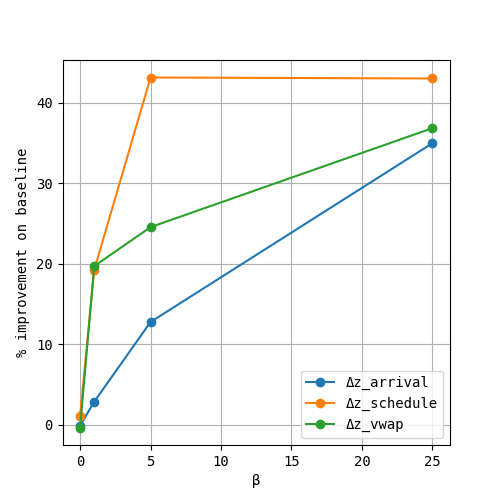}
	\caption{Performance improvement across slippage metrics (compared to spread crossing baseline) as we increase $\beta$.}
	\label{fig:beta_improvement}
\end{figure*}

\twocolumn

\end{document}